\documentclass[useAMS,usenatbib]{mnras}
\usepackage{natbib}
\usepackage{graphicx}

\title[Uncertainties on measuring  LSB galaxy distances]{IMF-induced intrinsic uncertainties on measuring   galaxy distances based on the number of giant stars: the case of the ultra-diffuse galaxy NGC 1052-DF2 }

\author[Zonoozi et al.]
{Akram Hasani Zonoozi$^{1,2}$\thanks{
E-mail:  \mbox{a.hasani@iasbs.ac.ir} (AHZ)},  Hosein Haghi $^{1}$, Pavel Kroupa$^{2, 3}$,
\\
$^{1}$Department of Physics, Institute for Advanced Studies in Basic Sciences (IASBS), PO Box 11365-9161, Zanjan, Iran\\
$^{2}$Helmholtz-Institut f\"ur Strahlen-und Kernphysik (HISKP), Universit\"at Bonn, Nussallee 14-16, D-53115 Bonn, Germany\\
$^{3}$Charles University in Prague, Faculty of Mathematics and Physics, Astronomical Institute, V Hole\v{s}ovi\v{c}k\'ach 2, CZ-180 00 \\
Praha 8, Czech Republic\\}

\begin{document}

\date{Accepted .... Received}

\pagerange{\pageref{firstpage}--\pageref{lastpage}} \pubyear{2013}

\maketitle

\label{firstpage}

\maketitle

\begin{abstract}

The surface brightness fluctuation (SBF) technique is one of the distance measurement methods that has been applied on the low surface brightness galaxy NGC 1052-DF2 yielding a distance of about 20 Mpc implying it to be a dark matter deficient galaxy.   We assume the number of giant stars above a given luminosity threshold to represent the SBF magnitude. The SBF magnitude depends on the distance, but this is degenerate with the star formation history (SFH). Using a stellar population synthesis model we calculate the number of giant stars for stellar populations with different galaxy-wide stellar initial mass functions (gwIMFs), ages, metallicities and SFHs.  If the gwIMF is the invariant canonical IMF, the 1$\sigma$ (3$\sigma$) uncertainty in colour allows a distance as low as 12 Mpc (8 Mpc).
If instead the true underlying gwIMF is the integrated galaxy-wide IMF (IGIMF) then overestimating distances for low-mass galaxies would be a natural result, allowing NGC 1052-DF2 to have a distance of 11  Mpc within the 1$\sigma$ colour uncertainty. Finally, we show that our main conclusion on the existence of a bias in the SBF distance estimation is not much affected by changing the luminosity lower limit for counting giant stars.

\end{abstract}


\section{INTODUCTION}\label{Sec:INTRO}


\Citealp{vanDokkum18} reported the  ultra-diffuse dwarf galaxy NGC 1052-DF2 (hereafter DF2)\footnote{For the discovery history of DF2 see \Citet{Trujillo18}.} with a total luminous mass of $M_{*}=2\times 10^8 M_\odot$  to lack dark matter. Its color $V_{606}-I_{814}=0.37\pm0.05$ (in the AB system), is similar to that of other ultra-diffuse galaxies and metal-poor globular clusters (GCs).  These authors used the line-of-sight velocities of 10 compact objects that are identified to be GCs  associated with DF2 as bright tracers of the gravitational potential. They obtained the dynamical mass of the galaxy to be $M_{dyn}<3.2\times 10^8 M_\odot$ which leaves  very little or no room for dark matter. With the obtained dynamical mass, the dark halo mass, $M_{halo}$, is of the order of the stellar mass which is 400 times lower than what is expected from the stellar-to-halo mass ratio \citep{Behroozi10}. They concluded that dark matter is not always coupled to baryonic matter on galactic scales. The $90\%$ confidence upper limit on the line-of-sight velocity dispersion of these objects is calculated to be less than $10.5$ km$s^{-1}$. However, estimating the velocity dispersion from only 10 tracers with large line of sight velocity errors leads to a significant uncertainty \citep{Martin18, Haghi19}.  Additionally, \citet{Lewis_2020}  uncovered evidence for rotation in the kinematics of DF2's GCs, which complicates the mass assessment in the galaxy. They found that  with a moderate inclination of the rotational component, the resultant dynamical mass-to-light ratio can exceed $(M/L)_{dyn}=10$.    

\Citet{vanDokkum19} announced the discovery of another dark matter free dwarf galaxy, NGC 1052-DF4 (hereafter DF4), in close proximity to DF2 with similar unusual properties. Based on the same procedure, from 7  GCs they derived a total enclosed dynamical mass within 7 kpc of $M_{dyn}=0.4^{+1.2}_{-0.3}\times 10^8 M_\odot$ and a total stellar mass of $M_{*}=(1.5 \pm0.4)\times 10^8 M_\odot$ and therefore inferred that it is the second dark matter deficient galaxy.

Using the surface brightness fluctuation (SBF) technique, the distance to DF2 and to DF4 is derived to be $19.0\pm1.7 $ Mpc and $19.9\pm2.8 $ Mpc, respectively. Due to the close projected distance in the sky, these galaxies are inferred to be  satellites of NGC 1052  and to lie at the same distance of about 20 Mpc \Citep{Tonry01}.  However, \cite{Blakeslee_2018}  carried out an independent analysis of the distance to DF2 using the SBF method and found a similar distance as \cite{vanDokkum18}, but showed that the result is only significant at 2$\sigma$, and therefore not definitive. Deep optical imaging of the system has revealed tidal tails around DF4 caused by its interaction with its neighboring galaxy NGC 1035 \citep{Montes_2020}.  \citet{Montes_2020} suggest that tidal stripping may remove a significant percentage of dark matter before affecting the stars of the galaxy as stars are more centrally concentrated than dark matter.

Moreover, the SBF distance \citep{vanDokkum18b},  the Planetary Nebula Luminosity Function distance estimate  of $20\,$ Mpc \Citep{Fensch18} and the non-detection of a gas component in DF2 \Citep{Chowdhury19}, suggest  this dwarf to be a member of the galaxy group NGC 1052.  Also, the Hubble Space Telescope should have been able to resolve the red-giant-branch stars if the galaxy was closer than $10\,$Mpc, unless this dwarf galaxy has a non-canonical stellar population.

Assuming a distance of $20$ Mpc, DF2 has a half-light radius of $2.2$ kpc, a central surface brightness of $\mu(V_{606,0})=24.4~mag~arcsec^{-2}$, a stellar mass of $M_{star}=2\times 10^8 M_\odot$, and has little dark matter. Therefore, this galaxy looks in size similar to the Milky Way but it is as faint as a dwarf galaxy\footnote{It is worth to mention that using the effective radius (which measures the light concentration of a  galaxy) for characterising galaxy size (which is related to the boundary of objects) has been criticized recently by \citet{Chamba_2020}. They showed that, while it is correct to say that UDGs and Milky Way like galaxies have comparable effective radii,  UDGs have radii within the range of dwarfs.}. In addition to the absence of dark matter, compared to the dwarfs of the Local Group, DF2 and DF4 are bigger and their population of GCs are unusually bright compared to typical GCs. Identifying DF4 in the same group with similar unusual properties, size, luminosity, an  unexpected luminous GC population with a low velocity dispersion,  \Citet{vanDokkum19} conclude that DF2 is not isolated and that a class of such large faint galaxies lacking  dark matter exists. The analysis of Illustris cosmological simulation by \cite{Haslbauer_2019} shows such galaxies to be extremely rare in the $\Lambda$CDM model. The detection of two such cases in close proximity thus constitutes a falsification of the $\Lambda$CDM model.

These conclusions and unusual properties depend on the adopted distance of 20 Mpc. If these two galaxies  lie at about 10 Mpc, then their stellar masses go down significantly and the ratio between dynamical and stellar mass increases such that DF2 and DF4 would become normal dark-matter containing dwarf galaxies. Moreover, for all known galaxies, the GC population universally peaks at $M_{\rm V}=-7.7\,$ \citep{Rejkuba12}, while that of DF2 peaks at $M_{\rm V} = -9.1$ for a distance of $D=20\,$Mpc. If DF2 were to lie at $D=8\,$Mpc, its GCs would appear normally bright and would have radii consistent with normal GCs. \citet{Trujillo18}  emphasized that using the empirical SBF relationship is wrong for a galaxy like DF2 (a dwarf), when the method was calibrated for massive ellipticals. They found a smaller distance of 13 Mpc for DF2 based on 5 different redshift independent distance indicators which is consistent with our earlier estimation based on the luminosity and  structural properties of DF2's GC population to be perfectly ordinary \citep{Kroupa18nat, Haghi19}.  Moreover, \citet{Monelli_2019} found that both DF4 and DF2 could be satellites of the spiral galaxy NGC1042 (which is at 13 Mpc). Adopting this distance all anomalies of the galaxy and its GC system disappear.




The SBF method \citep{Tonry88, Blakeslee_2012} is essentially based on the number of giants per unit projected surface area. The apparent magnitude and Poisson fluctuations across the image of the galaxy of this number can be used to tune it to a known system to get the absolute magnitude. If the fluctuations  then change, one can read off a new absolute magnitude, assuming a standard IMF. The SBF method can yield different results \citep{Cantiello_2003, Raimondo_2005, Carlsten_2019} because it relies on the number of giant stars per unit surface area which depends on the age and metallicity of the stellar population, the mass distribution, which also depends on the metallicity and star formation history (SFH,  \citealp{Kroupa13, Yan2017, Jerabkova18, Zonoozi19}).
 
Therefore, the main challenge regarding the discovery of dark matter deficient galaxies is the uncertainty in distance that comes from the uncertainty in the SFH and the presence or absence of a conspicuous asymptotic giant branch (AGB) population that might affect distance measurements both for the SBF and tip of the red-giant branch (TRGB) methods. While it is true that even the TRGB method yielded disputed results for the distance to DF2 \citep{Trujillo18, Cohen_2018, vanDokkum18c},  most recently, \citet{Danieli_2020} obtained deeper data of  DF4 with the HST (8 orbits in the F814W filter and 4 orbits in the F606W filter) and measured a distance of $20 \pm 1.6$ Mpc for DF4 using the TRGB method. Their observations pointed to the outer parts of the galaxy, where blending is reduced. Theoretically an intrinsic difference in the SBF magnitude due to  a difference in the age and metallicity of the stellar population is thus expected \citep{Mei05,Blakeslee09,Blakeslee10,Jensen15,Cantiello 18}.

Here,  we study the influence of the galaxy-wide IMF (gwIMF) and of the SFH on the number of bright giant stars.  We emphasise that this contribution is not meant to be a full-scale and detailed analysis of the SBF method. The primary aim is to point out how the distance measure, $D \propto 1/\sqrt{N}$, where $D$ is the distance and $N$ the number of stars which define the surface brightness fluctuation (usually being giants), leads to different results on $D$ if a plausible, different-than-the-canonical gwIMF is applied. Thus, we aim to investigate by how much $\sqrt{N}$ changes for the galaxies at a fixed same distance but with different SFH and metallicity which define the shape of the gwIMF.

Section 2 presents a brief introduction to the SBF method for distance determination. In Section 3 the here adopted method of calculating the gwIMF is summarized. In Section 4, we discuss our results and assess how the star formation rate (SFR) and the gwIMF of a galaxy affects the distance measurement based on the SBF technique. Finally, in Section 4 we present our conclusions.

\section{The surface brightness fluctuation method}

The interpretation of the essential properties of a galaxy, such as its luminosity, radius, star formation rate (SFR), and visible mass depends on its distance. Therefore, a large effort is invested in improving the accuracy of distance measurement methods.
Determining galaxy distances from the amplitude of their fluctuations in surface brightness \Citep{Tonry88} is one of the useful distance indicators. The SBF method is based on the fundamental fact that the number of stars per resolution element has a statistical fluctuation equal to the square root of the average number of stars. Assuming a galaxy at a determined distance $d=d_0$, in the CCD image of a certain region of that galaxy, there are on average $N_0$ stars imaged in each pixel. Therefore, there will be a $100\sqrt {N_0}/N_0$ per cent  pixel-to-pixel fluctuation of that number and correspondingly a $100\sqrt {N_0}/N_0$ per cent pixel-to-pixel fluctuation of the measured flux. Now assume the same galaxies at a distance $d=kd_0$. In the same region as before, there will be $k^2N_0$ stars per pixel, since the metric area sampled by one pixel increases by a factor of $k^2$. The corresponding pixel-to-pixel fluctuations decrease to $100\sqrt {k^2N_0}/(k^2N_0)$ per cent of the mean flux, being lowered by a factor of $1/k$. That is, the amplitude of the surface brightness fluctuations is inversely proportional to the distance and can therefore be used as a distance indicator. The more distant galaxies look smoother, while nearby ones appear bumpy. This is because the number of unresolved stars per CCD pixel increases with distance. As a consequence, while  the galaxy surface brightness remains independent of distance, the surface brightness fluctuation decreases inversely with the distance, $d$. 
Note that the SFB method is most useful in the regime where the galaxy is partially resolved. If the galaxy is too close, the individual stars will contribute flux to different pixels, while if the galaxy is too far, the fluctuations become too small to be detected.

The SBF amplitude depends on the stellar population properties and is most sensitive to the properties of stars in the brightest evolutionary phases within the observational passband. In the optical and near-IR these will be red giants (either RGB or AGB), but in star forming galaxies there may also be a contribution from red supergiants. For early-type stellar populations the fluctuations are dominated by the variation in the number of stars on the red giant branch in each pixel. Some theoretical models of SBFs in the optical and near-infrared bands that explore effects due to the variation of the IMF and of the stellar color-temperature relation  adopting the Teramo-SPoT simple stellar population (SSP) models have been presented by \cite{Cantiello_2003, Raimondo_2005} and \cite{Raimondo_2009}.  They provided theoretical SBF amplitudes for single-burst stellar populations of young and intermediate age and different metallicities. Specifically, \cite{Cantiello_2003} evaluated the contribution of the lowest mass stars to the SBFs by adopting an IMF starting from different low mass limits as well as by arbitrary changing the slope of the IMF and showed that the effect on the SBF values is rather small as the total luminosity of the stellar population does not change significantly.

The SBF method is used and calibrated on different types of galaxies from ellipticals to early-type dwarfs and LSB galaxies \citep{Jerjen_1998, Jerjen_2000, Rekola_2005, Dunn_2006, Carlsten_2019}.
Particularly, \cite{Carlsten_2019} presented a new calibration of the SBF method to measure distances to very low surface brightness galaxies (similar to DF2). This work tied the SBF calibration to the tip of the red giant branch distance determination method, which is well established in the literature and they acknowledge that in spite that the intrinsic scatter increases for blue galaxies, the rms of the calibration still makes the SBF measurements useful, albeit not as precise as the red giant branch tip (TRGB) method. A review of the SBF method and its use for primary and secondary distance determination can be found in \cite{Blakeslee_2012}.

Assuming a composite stellar population for a galaxy, the SBF magnitude can be sensitive to the gwIMF, metallicity, age and SFR.  Above, there are known correlations of the SBF magnitude with colour (metallicity) and star formation history that need to be taken into account for determination of distance. We use the IGIMF theory \citep{Kroupa03, Kroupa13} to construct the present-day gwIMF of low mass galaxies like DF2. Based on the IGIMF theory, which is formulated to yield consistency with observed resolved stellar populations, a larger SFR leads to a more top-heavy gwIMF which means that the galaxy will have more luminous stars.  This comes about because galaxies with a higher SFR also form more massive embedded star clusters which have, above a certain mass, a top-heavy stellar IMF (Eq. \ref{alpha3} below). Here we distinguish between the stellar IMF in an embedded cluster and the gwIMF which is the sum of all IMFs in all embedded clusters (see also \citealt{Jerabkova18} for more detailed discussion).  Depending on the age of the  model and it's SFH, some of these luminous stars will contribute to the SBF magnitude.

\section{Methods}

\subsection{The Galaxy-wide IMF}


The canonical stellar IMF is the distribution of stars formed together in a molecular cloud core, i.e., in an embedded cluster,  with upper and lower limits of 100 $M_\odot$ and 0.08 $M_\odot$,  respectively, and is defined as  a two-part power-law function \citep{Kroupa01}, 

\begin{eqnarray}
\xi (m) \propto m^{ -\alpha}: \left\{
\begin{array}{ll}
\alpha _1 = 1.35 \hspace{0.25 cm} ,\hspace{0.25 cm} 0.08 \leq \frac{m}{M_\odot} < 0.5,\\
\alpha _2 = 2.35 \hspace{0.25 cm} ,\hspace{0.25 cm} 0.5 <\frac{m}{M_\odot} \leq 100.\\
\end{array}
\right.
\end{eqnarray}
where, the number of stars in the mass interval $m$ to $m + dm$ is $dN=\xi (m)dm$. 


Observations indicate that the IMF may depend on the star formation environment (cloud density and metallicity)  becoming top-heavy under extreme starburst conditions \citep{Dabringhausen09, Dabringhausen10, Dabringhausen12, Marks12, Banerjee12, Kroupa13, Schneider18, Kalari18, Jerabkova17}. The data suggest that the IMF becomes less top-heavy with increasing cluster metallicity and decreasing density. The need for an IMF variation has been also put forward to interpret the evidence coming from mass-to-light ratios estimated through integrated light analysis of GCs in M31 which show an inverse trend with metallicity \citep{Zonoozi16, Haghi17}, and the fraction of low-mass X-ray binaries in Virgo GCs and ultra compact dwarf galaxies \citep{Dabringhausen12}. 


The correlation between the metallicity and molecular cloud core (embedded cluster) density, $\rho_{cl}$ (in gas and stars),  and the power-law index of the IMF in an embedded cluster for stars more massive than 1 $M_{\odot}$, $\alpha_3$, can be written as \Citep{Marks12}:

\begin{eqnarray}
\alpha_3=
\left\{
       \begin{array}{ll}
          2.3    & x<-0.87,  \\
          -0.41x+1.94     & x>-0.87,
 \end{array}
          \right. \label{alpha3}
\end{eqnarray}
where $x=-0.14[Fe/H]+0.99\log_{10}(\rho_{cl}/10^6M_\odot pc^{-3} )$.  The density $\rho_{\rm cl} = 3\,M_{\rm ecl} / (4\, \pi\, r_{\rm h}^3)$, where the half-mass radius is given by $r_{\rm h}/{\rm pc}= 0.1(M_{\rm ecl}/M_{\odot})^{0.13}$ \citep{Marks_2012} and a star formation efficiency of 33 per cent is assumed \citep{Lada03, Megeath16}.

Assuming that all stars form in embedded clusters \citep{Kroupa_1995a, Kroupa_1995b, Porras2003, Lada03, Kroupa05, Megeath16}, and adding the IMFs of all clusters formed in a star formation epoch of the galaxy, \citet{Kroupa03} formulated the integrated galaxy IMF (IGIMF) theory to quantify the gwIMF. Based on the IGIMF theory \citep{Yan2017, Jerabkova18}, the gwIMF is more top-heavy in massive galaxies with a high SFR as is observed \citep{ Hoversten08, Lee09, Meurer09, Habergham10, Gunawardhana11,  Zhang18, Hopkins18}, while it becomes top-light (deficit of massive stars) in low mass galaxies which have a low-level of star formation activity \citep{Ubeda07, Lee09, Watts18, Yan2017}. The application of the IGIMF theory to the chemical evolution of ultra-faint dwarf galaxies yields excellent agreement with observations and demonstrates how the IMF can be constrained in extreme environments \citep{Yan2020}.

A detailed description of the IGIMF concepts and of the nomenclature and  full IGIMF grid for galaxies with different SFR and metallicity  is provided by \citet{Jerabkova18}.  Old, dormant galaxies also show significant variations of their stellar populations: elliptical galaxies may be dominated by very low mass stars \citep{vanDokkum10,vanDokkum11, Conroy17}, while ultra-faint dwarf galaxies may have a deficit of low mass stars  \citep{Geha13, Gennaro18} when compared to the canonical stellar population.  

The IGIMF is the sum of all the stars in all embedded clusters formed over a time interval $\delta t$ (the star formation epoch)  and  is calculated as the integral over the  embedded star cluster MF, $\xi_{ecl}(M) $  \citep{Yan2017, Jerabkova18},

\begin{eqnarray}
\xi_{IGIMF}(m,\psi(t))=\,\,\,\,\,\,\,\,\,\,\,\,\,\,\,\,\,\,\,\,\,\,\,\,\,\,\,\,\,\,\,\,\,\,\,\,\,\,\,\,\,\,\,\,\,\,\,\,\,\,\,\,\,\,\,\,\,\,\,\,\,\,\,\,\,\,\,\,\,\,\,\,\,  \nonumber\\
\int_{M_{ecl,min}}^{M_{ecl,max}(\psi(t))}\xi(m\leq m_{max})\xi_{ecl} (M_{ecl},\psi(t)) dM_{ecl}. \label{IGIMF}
\end{eqnarray}
The IMF, $\xi(m)$,  depends on the  metellicity $[Fe/H]$ and density through $\alpha_3$ (Eq. \ref{alpha3}) and through the most-massive-star-$M_{ecl}$ relation, $m_{max}(M_{ecl})$ \citep{Weidner_2013}. $M_{ecl,max}$ is the maximum mass of the embedded cluster that can form in a star formation epoch and depends on star formation rate (SFR), $\psi(t)$ \citep{Weidner_2004}. $M_{ecl,min}=5 M_{\odot}$ is the minimum mass of an embedded cluster as observed in the Taurus-Auriga star forming region (e.g. \citealt{Briceno1998, Kroupa2003, Joncour2018}). The star formation epoch, $\delta t=10$ Myr, is the timescale on which the interstellar medium forms a complete population of stars in embedded clusters and is discussed in \cite{Schulz2015} and \cite{Jerabkova18}.

The IGIMF theory is used here to study the influence of a systematically varying gwIMF on the number of giant stars.



\subsection{The SFH, metallicity and stellar population model}

The SFH,  age and metallicity (and thus also the colour) of the stellar population of a galaxy is likely to have an important impact on the distance measurement based on the SBF magnitude. 

We assume that for distant galaxies like DF2, the number of giant stars is  related to the SBF magnitude. Based on this assumption and using a stellar population synthesis model we determine the number of giant stars and the color of the stellar system with different starting and truncating times of star formation.
For each model the integrated $V_{606}-I_{814}$ color is calculated using the stellar evolution tracks from the Padova group \Citep{Marigo07, Marigo08} and the semi empirical BaSeL3.1 library \Citep{Lejeune97,Lejeune98,Westera02} as well as an empirically calibrated  theoretical stellar spectral library. The gwIMF is adopted to be either canonical or the IGIMF.    

For all models we assume that the SFR is constant and the metallicity is $ [Fe/H]=-1$  corresponding to the measured metallicity for DF2  \Citep{vanDokkum18}. However in order to consider the influence of metallicity on the color and number of giant stars and consequently on the SBF magnitude, two more series of models with [Fe/H]=-1.3 and [Fe/H]=-0.7 are calculated. The total created stellar mass is assumed to be $M_{tot}=3 \times10^8 M_\odot$.
This gives  a present-day mass roughly in the range of  $1.5-2\times10^8 M_\odot$, which is in agreement with the reported stellar mass for DF2 and DF4 if these galaxies are located at 20 Mpc. Therefore, the SFR is  $SFR=M_{tot}/(T_{trunc}-T_{start})~\frac{M\odot}{Myr}$. Here, $T_{start}$ and $T_{trunc}$ are the start- and end-time of the SFH, respectively.

\section{Results}

In this section,  we address some intrinsic uncertainties that may lead to an error in the distance measurement based on the SBF method.  To see to what extent the number of giants depends on the gwIMF and age of a low surface brightness galaxy like DF2, we  calculated a series of stellar population models with  a constant SFR and the same metallicity, but a different start and truncation time of star formation. Depending on the model, the number of giant stars is computed assuming the gwIMF is the   canonical IMF or the IGIMF.

\begin{figure}
 \begin{center}
 \includegraphics[width=90mm,height=75mm]{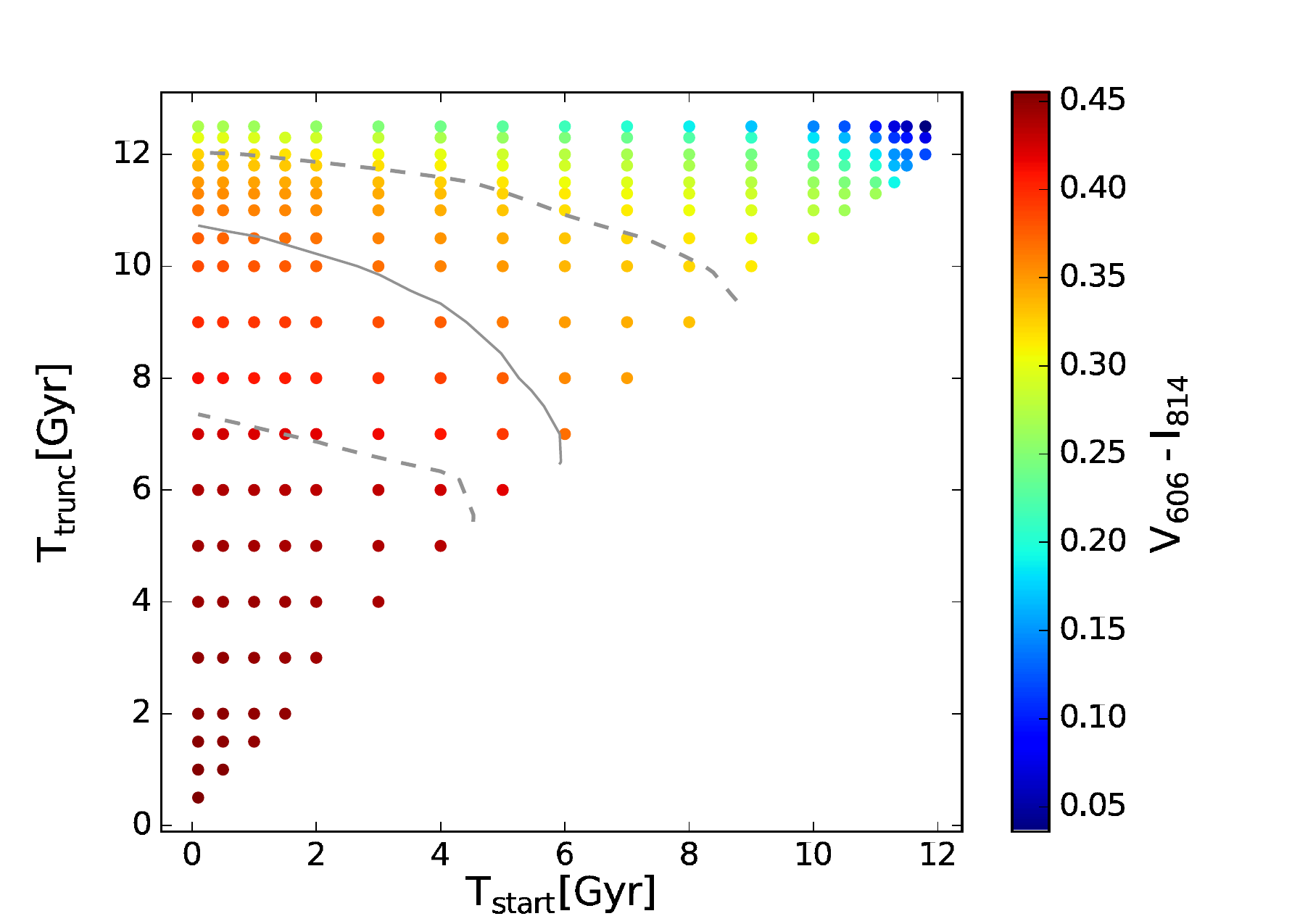}
 \caption{The $V_{606}-I_{814}$ color of the stellar population with different start and truncation times of star formation. 
 For all models the gwIMF is canonical and   $[Fe/H]=-1$ is adopted. Solid line shows DF2's observed color, $V_{606}-I_{814}=0.37\pm0.05$ \Citep{vanDokkum18} and dashed lines indicate its 1$\sigma$ uncertainty.}
 \label{V-I}
 \end{center}
 \end{figure}

\begin{figure}
 \begin{center}
 \includegraphics[width=85mm,height=65mm]{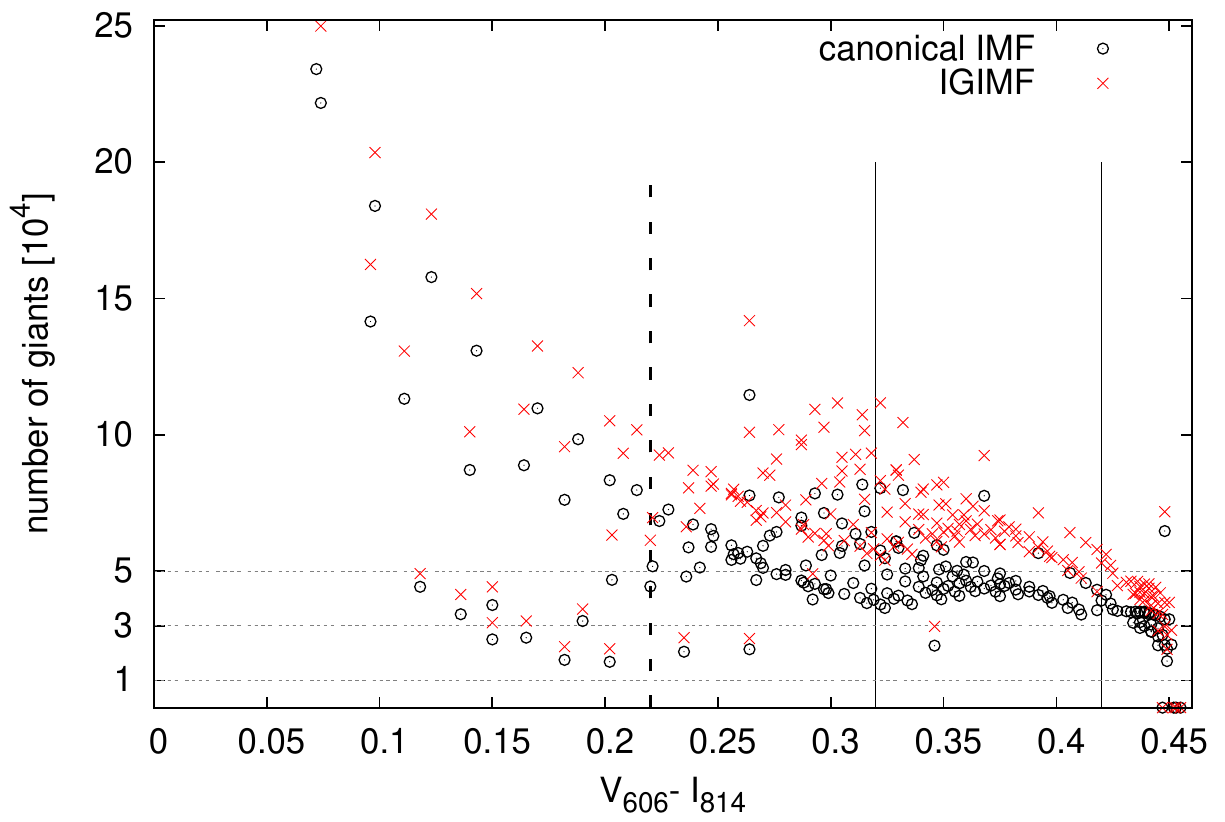}

 \caption{ Number of giant stars with luminosity above $\rm log_{10}(L/L_{\odot})_{bol} = 2.8$  versus $V_{606}-I_{814}$ color for models with different start and truncation time for star formation. Each point represents a galaxy with individual start/truncation time (shown in Figure \ref{V-I}).  Red crosses are for models assuming the gwIMF is the IGIMF, while black circles assume it to be the canonical IMF.  The vertical solid lines indicate the left and right limit of the  1$\sigma$ uncertainty in the measured color with the central value of $V_{606}-I_{814}=0.37\pm0.05$ for DF2 \Citep{vanDokkum18}. The vertical dashed line shows the left limit of the 3$\sigma$ colour uncertainty. Horizontal lines are drawn to better identify the location of points.}
 \label{Ng-color}
 \end{center}
 \end{figure}

\subsection{The effect of the SFH on the distance of DF2}

As illustrated in Figure \ref{V-I}, comparing the color  of the composite stellar population models with the observed $V_{606}-I_{814}$ color of DF2, $T_{start}$ and $T_{trunc}$  can be constrained for this galaxy. As the  age of the Universe is assumed to be about $13.5 $ Gyr, the age of each model is $T_{age}=12.5-T_{start}$ Gyr, i.e. we assume the galaxy started to form at the earliest 1 Gyr after the Big Bang. Based on these assumptions  we found that the lower and upper limit of the age of DF2 could be in the range of roughly $3.5-12.5 $ Gyr and its star formation occurred  up to 0.5 to 6.5 Gyr ago. 

By a full spectral fitting analysis \Citet{Fensch18} and \citet{Ruiz-Lara_2019} reported that the stellar population of DF2 is $8.9\pm1.5$ and $8.7\pm0.7$ Gyr old, respectively. This is comparable with the age that we obtain based on the color for single stellar population models with the best-fitting age of $6.5^{+2}_{-3} $ Gyr (Figure \ref{V-I}).  Moreover, our estimate is similar to the most likely age of 5.4 Gyr found  by \Citet{Trujillo18} using the SED fitting method.





Figure \ref{Ng-color} (black circles) shows the number of giant stars with luminosity above  $\rm log_{10}(L/L_{\odot})_{bol} \approx 2.8$ (which is a typical lower luminosity limit of giant stars) versus the $V_{606}-I_{814}$ color for models assuming the gwIMF is the canonical IMF and IGIMF for different $T_{start}$ and $T_{trunc}$. Each point represents a galaxy with individual start/truncation time (shown in Figure \ref{V-I}).  
As illustrated in Figure \ref{Ng-color},  the number of giant stars increases for a bluer $V_{606}-I_{814}$ colour.
The observed range of the color of DF2 is indicated by the vertical lines. The number of bright stars for different models varies from about $3\times 10^{4}$ to $8\times 10^{4}$ (for the case of the canonical IMF) in the observed color uncertainty which implies a difference in distance measurement by a factor of  about $\sqrt{\frac{8}{3}}$. 
This means that the distance estimate ranges from 12 to 20 Mpc  for the 1$\sigma$ colour uncertainty and may reach even smaller distances by a factor of  $\sqrt{\frac{12}{2}}$, reaching 8 Mpc at the 3$\sigma$ blue extreme,  if the canonical IMF is assumed for the stellar population.  The different branches in Figure \ref{Ng-color} correspond to different $T_{trunc}$ for populations that had  star formation recently. As these models have a larger number of luminous giants which is very sensitive to the adopted $T_{trunc}$, smaller bins of $T_{trunc}$ would be needed to cover all possibilities of color and number of giants in between the branches. Note that the total mass of the stellar population, $M_{tot}$, matches the observed stellar mass of $3\times 10^{8} M_{\odot}$, which, combined with the relatively low limiting luminosity, accounts for the observed number of giants in HST imaging data, while the real galaxy has in total $> 10^{5}$ giants.

\subsubsection{Different luminosity limits for counting giant stars}

\begin{figure}
  \begin{center}
   \includegraphics[width=85mm,height=65mm]{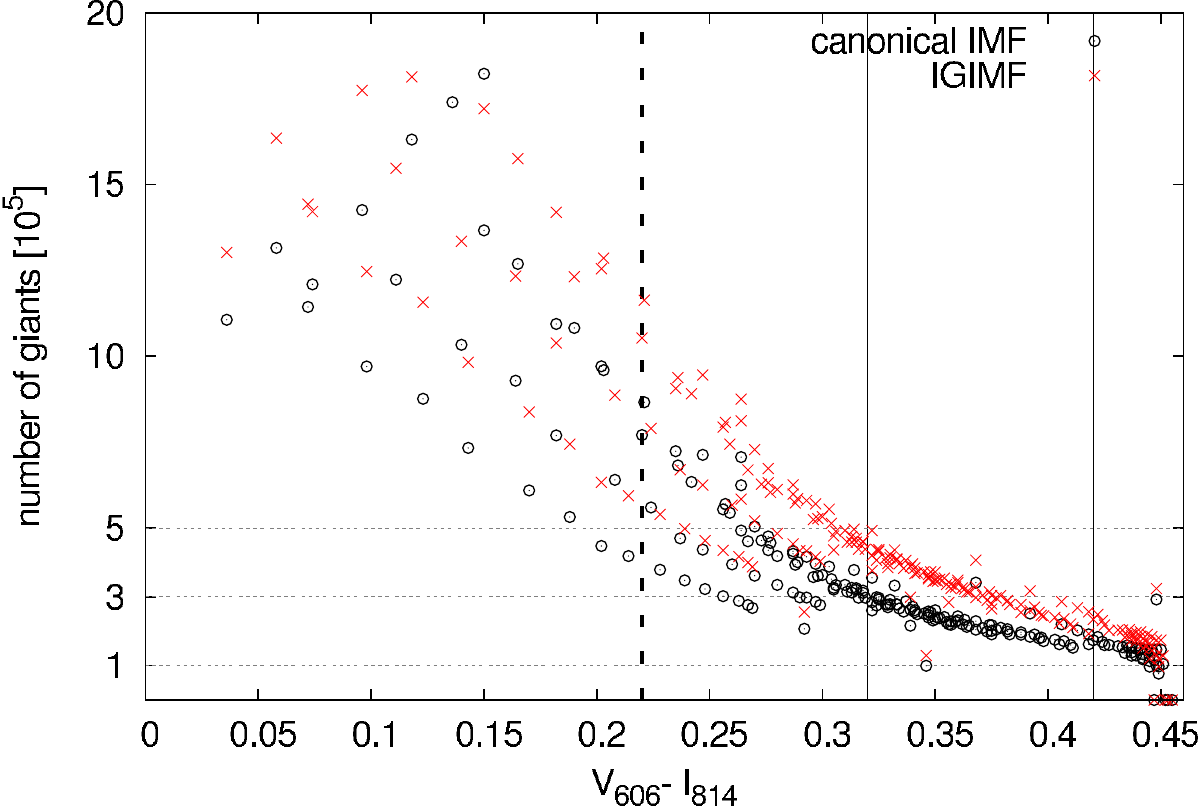}
  \caption{ Same as Figure \ref{Ng-color} but for the fainter value of $\rm log_{10}(L/L_{\odot})_{bol} = 2$ as the lower luminosity limit for counting the number of giant stars. For each symbol different branches from bottom to top correspond, respectively,  to different truncation times for star formation that occurred 100, 200, and 500 Myr ago.}
  \label{Ng-color-L100-1}
  \end{center}
  \end{figure}

The above mentioned choice of    $\rm log_{10}(L/L_{\odot})_{bol} \approx 2.8$ as the luminosity limit for counting the number of giant stars is based on this value being a typical luminosity of giants. In order to show  how the results change considering a fainter cutoff for the luminosity we repeat our calculation for $L_{bol}=100~L_{\odot,bol}$.  Figure \ref{Ng-color-L100-1}  shows the number of giant stars with luminosity above $\rm log_{10}(L/L_{\odot})_{bol} = 2$ versus the $V_{606}-I_{814}$ color for different $T_{start}$ and $T_{trunc}$. The number of bright stars for different models varies from about $1.5\times 10^{5}$ to $3\times 10^{5}$ (for the case of the canonical IMF) in the observed color uncertainty which implies a difference in distance measurement by a factor of  about $\sqrt{\frac{3}{1.5}}$. 
This means that the distance estimate ranges from 14 to 20 Mpc  for the 1$\sigma$ colour uncertainty and may reach even smaller distances by a factor of $\sqrt{\frac{8}{1.5}}$, reaching 8.7 Mpc at the 3$\sigma$ blue extreme,  if the canonical IMF is assumed for the stellar population.

\begin{figure}
  \begin{center}
   \includegraphics[width=85mm,height=65mm]{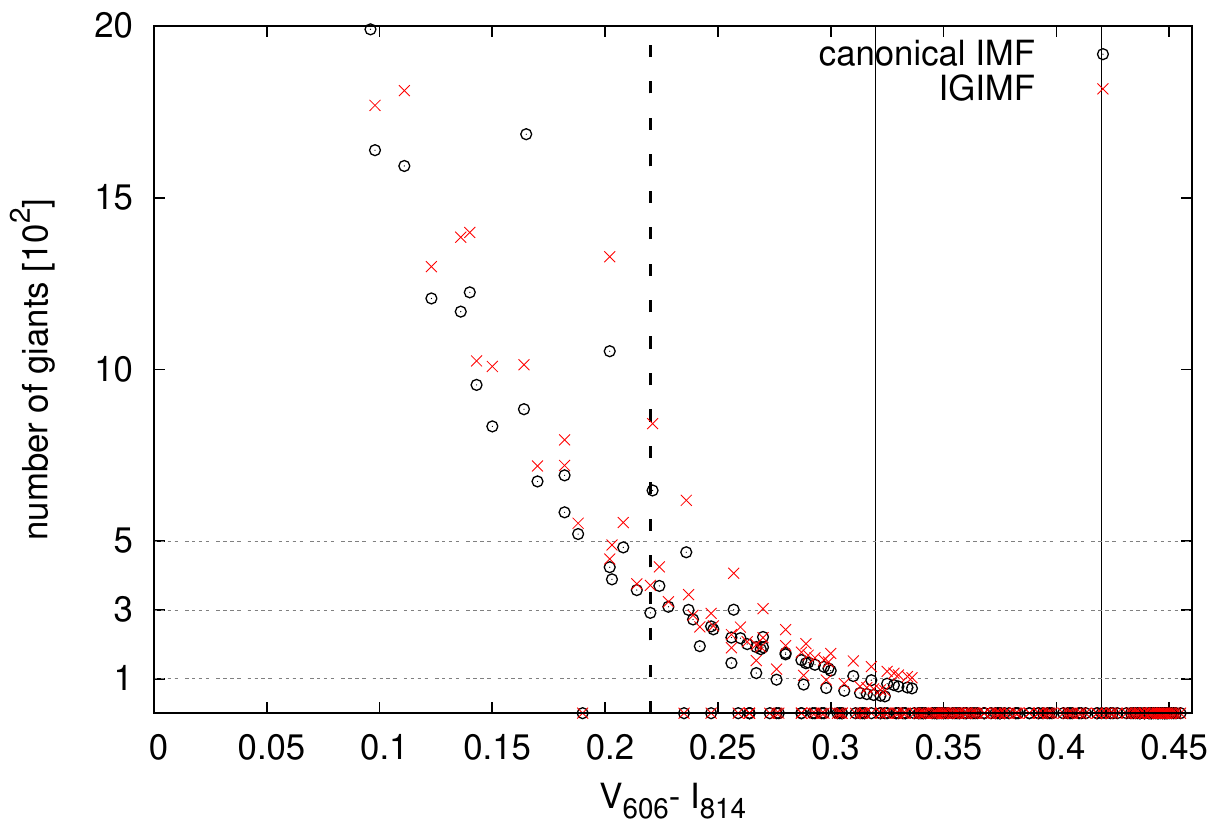}
  \caption{ Same as Figure \ref{Ng-color} but for the brighter value of $\rm log_{10}(L/L_{\odot})_{bol} = 3.5$ as the lower luminosity limit for counting the number of giant stars. The points with zero number of giants are related to old models with low truncating times of star formation so that these models do not have luminous giants with $\rm log_{10}(L/L_{\odot})_{bol} \geq 3.5$.}
  \label{Ng-color-L35}
  \end{center}
  \end{figure}

The mean luminosity of the giants however depends on the age of the population. The giant stars that contribute most to the SBF fluctuations are those found very close to the tip of the RGB or in the upper AGB. In an old ($>$4 Gyr) stellar population the TRGB is found at $\rm log_{10}(L/L_{\odot})_{bol} \approx 3.3-3.5$ depending on the metallicity of the population.  In order to explore how the number of giant stars depends on color for different start/end times of the star formation and on the brighter limit for the luminosity threshold, we repeat our calculation assuming a luminosity limit of $\rm log_{10}(L/L_{\odot})_{bol} \approx 3.5$  for the number of giants we count. The result is shown in Figure  \ref{Ng-color-L35}. As can be seen, the selected luminosity limit is so bright that it is appropriate only for younger models. Therefore, for redder colours, corresponding to older models whose star formation period was much earlier, there are no giants observed at all (shown as points with a zero number of giants in Figure  \ref{Ng-color-L35}). That means also that SBFs would not be observed at this bright limit in DF2. Between the bluer solid vertical line and the dashed vertical line in Figure \ref{Ng-color-L35} (3$\sigma$ uncertainty limits on DF2's colour) the number of giant stars increases by a factor of 15 (for the case of the canonical IMF).



For models where the star formation ended less than $1-2$ Gyr ago the contribution of AGB stars to SBFs starts to matter. While for the bluer/younger models the limit of $\rm log_{10}(L/L_{\odot})_{bol} = 3.5$ may be relevant, in order to include the entire colour uncertainty range (and thus the relevant star formation histories) one should include also giant stars fainter than the TRGB (i.e. in the range of about $\rm log_{10}(L/L_{\odot})_{bol} \approx 2.8-3.4$) as we have considered in this paper.




%



 \begin{figure}
  \begin{center}
  \includegraphics[width=85mm,height=65mm]{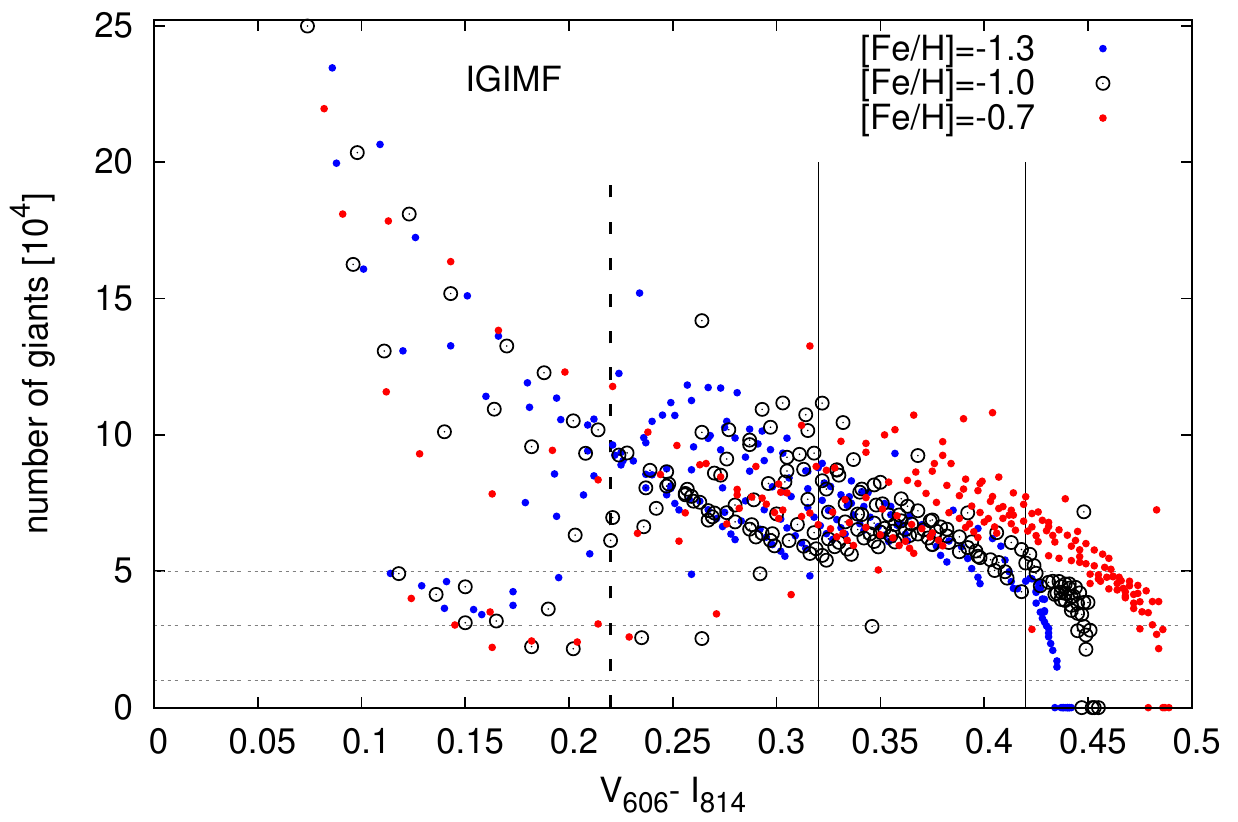}
  \includegraphics[width=85mm,height=65mm]{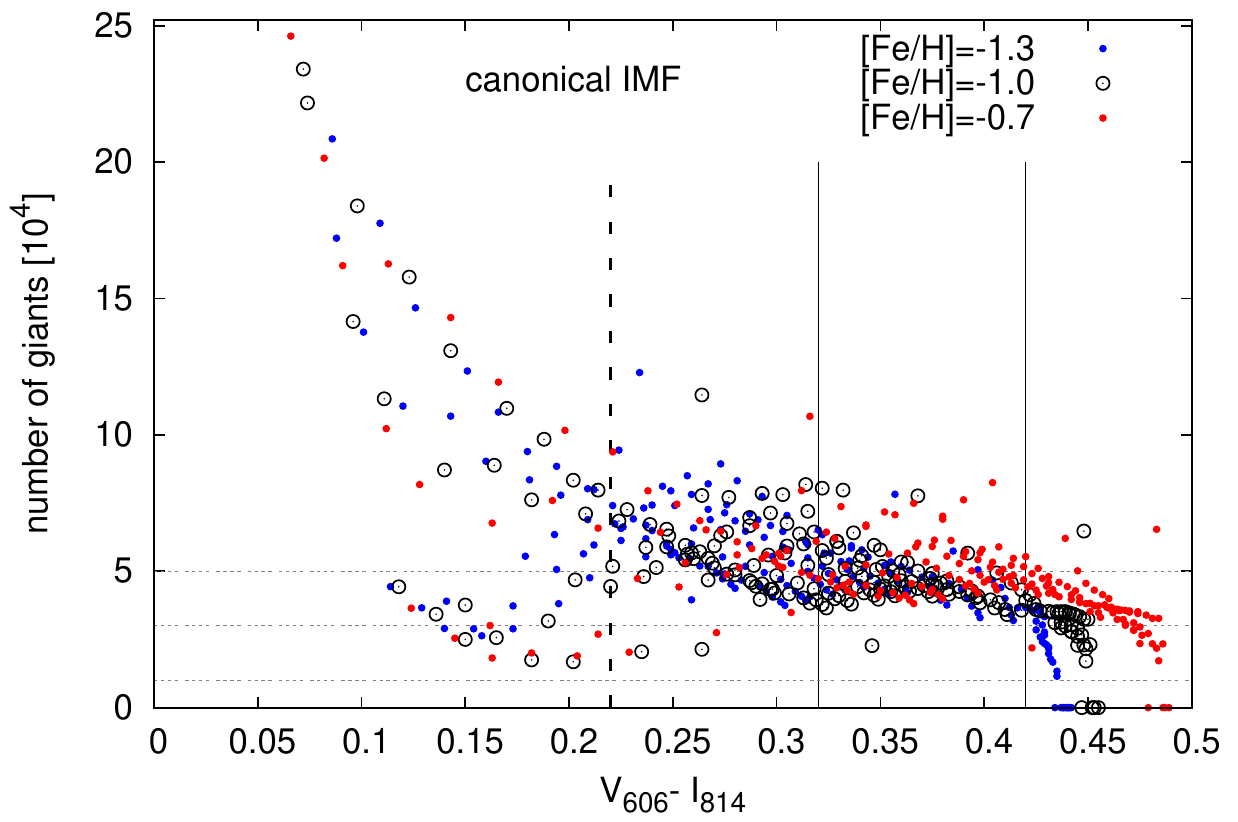}
  \caption{ Same as Figure \ref{Ng-color} but for different metallicities, assuming the gwIMF is the IGIMF (upper panel) and canonical IMF (lower panel). }
  \label{Ng-color-1}
  \end{center}
  \end{figure}
  
\begin{figure}
 \begin{center}
 \includegraphics[width=95mm,height=75mm]{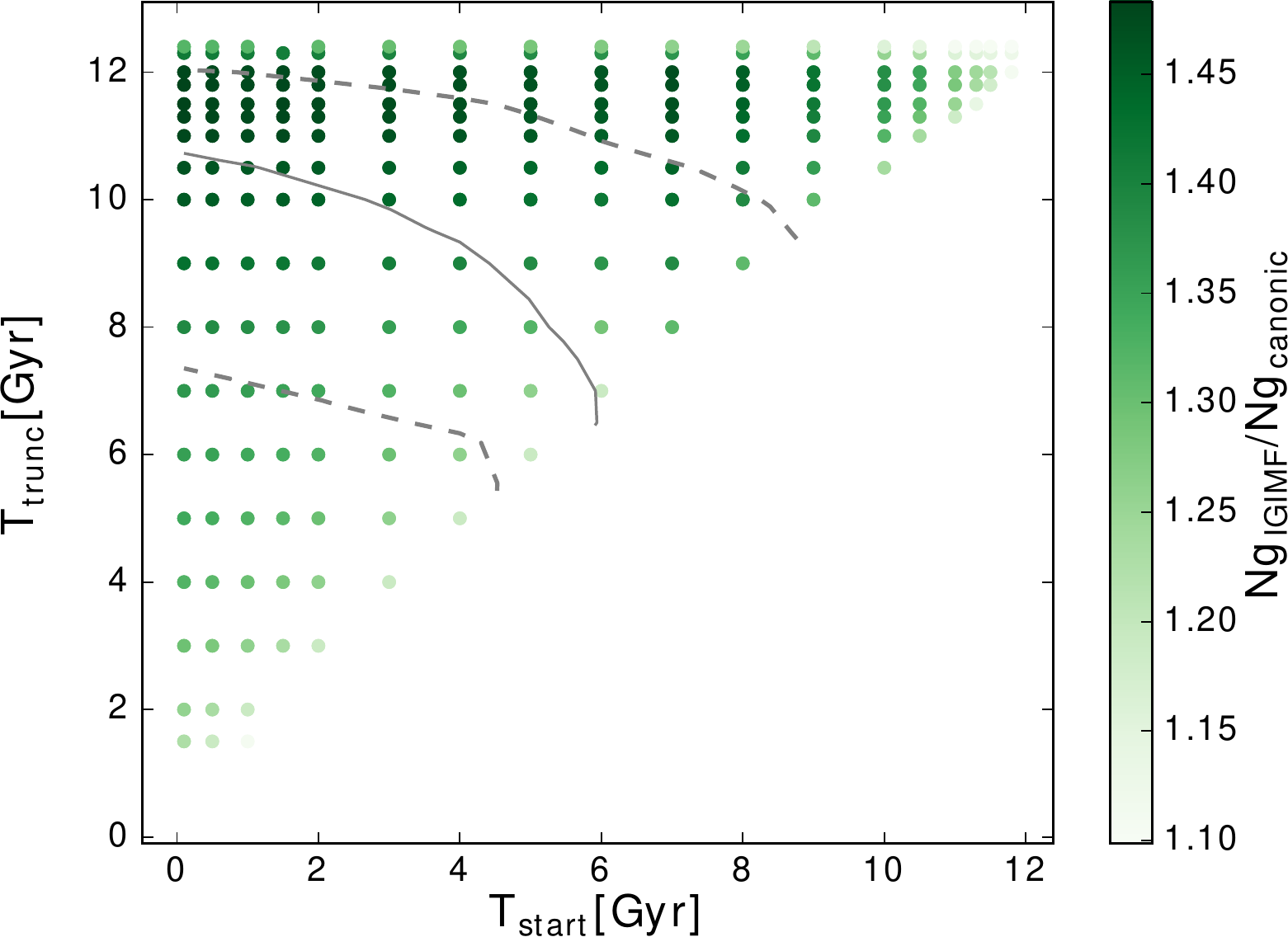}
 \caption{The influence of the stellar IMF on the number of giant stars, $N_g$. Assuming the IGIMF and canonical IMF, the number of giant stars with luminosity above $\rm log_{10}(L/L_{\odot})_{bol} = 2.8$ is computed for each model and their ratio, $N_{g, IGIMF}/N_{r, canonical}$, is plotted in the two-parameter space of $T_{start}$-$T_{trunc}$. The solid line shows models  with the $V_{606}-I_{814}$  color being equal to the observed value and the dashed lines show the 1$\sigma$ uncertainty range.}
 \label{IG-IM}
 \end{center}
 \end{figure}

\subsection{The effect of a non-canonical gwIMF on the distance of DF2}


We use the IGIMF theory to construct the  present-day gwIMF of low mass galaxies like DF2 to see how  the number of giant stars (and by implication the distance measurement based on the SBF magnitude) is affected in comparison to assuming a canonical IMF. The number of bright giant stars of each model is compared in Figure \ref{Ng-color} for the cases that  the gwIMF is the invariant canonical IMF and the IGIMF. As can be seen, the number of giant stars for different models varies from about $4(3)\times 10^{4}$ to $12(15)\times 10^{4}$ (for the case of the IGIMF) in the 1(3)$\sigma$ color uncertainty which implies the distance estimate ranges from 11.5(9) to 20 Mpc  for the 1(3)$\sigma$ colour uncertainty.

Moreover, in this section, the effect of metallicity on the number of giant stars in the context of the IGIMF theory and canonical IMF is shown in Figure \ref{Ng-color-1}. As can be seen, the metallicity has an important influence on the number of giants. For example, by increasing the metallicity from $[Fe/H]=-1.3$ to $-0.7$ the number of giants increases by a factor of about 2  at the color   $V_{606}-I_{814}=0.37$ in both the IGIMF and  the canonical IMF cases.   

In the upper panel of Figure \ref{Ng-color-1}, it can be seen that the combination of the effect of the intrinsic uncertainties on the SFH (age and metallicity) and the IGIMF gives a wider range of the number of giant stars from $N_{1}=4(2)\times10^4$ to $N_{2}=13(15)\times10^4$ within the 1(3)$\sigma$ color uncertainty. This could reduce the distance estimate from 20 Mpc to 11 Mpc (1$\sigma$ uncertainty) and even down to 7 Mpc (3$\sigma$ uncertainty).

Figure \ref{IG-IM} shows the ratio of giant stars for models constructed based on the IGIMF to that using the canonical  IMF. Different models have different SFRs  that lead them to have different gwIMFs according to the IGIMF theory.  The ratio of the number of giant stars in stellar population models constructed based on the IGIMF theory to models adopting the canonical IMF varies from 1.15 up to 1.5. This means if the real gwIMF of galaxies follows the IGIMF theory instead of the universal canonical and invariant IMF then galaxies like DF2 with the same color, metallicity and mass will have more luminous giant stars than what we expect from the canonical IMF.  Therefore, this effect alone can imply a change in the distance estimate from 20 Mpc to 17 Mpc, as $D^2 \propto N$. 

\begin{figure}
 \begin{center}
 \includegraphics[width=80mm,height=75mm]{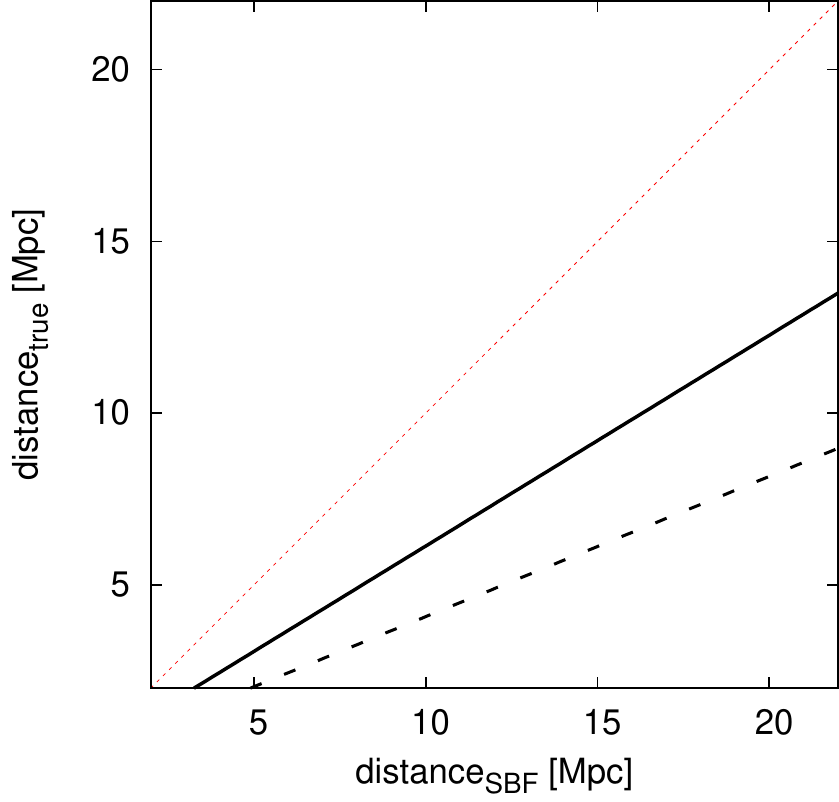}
 \caption{The assumed true distance vs. the SBF distance (which is used in \citealt{vanDokkum18}) as given by the  number of giant stars. The dotted line shows unity. Solid (dashed) line shows the maximum SBF distance within 1(3)$\sigma$ colour uncertainty. Thus, a SBF distance of 20 Mpc can correspond to a true distance down to 7 Mpc within the current 3$\sigma$ uncertainty on the colour of DF2, if the DF2-like galaxy is described by the IGIMF theory rather than an invariant canonical IMF.  }
 \label{dSBF}
 \end{center}
 \end{figure}

In Figure \ref{dSBF}, we demonstrate how much the estimated distance based on  the number of giants (and thus as based on the SBF method)  could be different from the assumed true distance. The maximum SBF distance that can be obtained is calculated including the effect of both color uncertainty and different IMFs as 

\begin{equation}
    \frac{D_{SBF, max}}{D_{true}}= \sqrt{\frac{N_{2}}{N_{1}}}.
\end{equation} 
The maximum "SBF distance" in the 1$\sigma$ (solid line) and 3$\sigma$ (dashed line) colour uncertainties is plotted for dwarfs located at different assumed true distances.

Since DF2 and DF4 have very similar properties in terms of having a very low mass density, metallicity and past SFR such that they probably have a similar formation history, we expect DF4's distance measurement as based on  the SBF method to suffer from the same bias.

\section{Conclusion}





We have investigated the effect of the intrinsic uncertainty of the SBF method on  measuring the distance of ultra-diffuse dwarf galaxies like DF2. We determined to what extent this distance measurement  can be affected by the SFH and a varying gwIMF using the IGIMF theory. We emphasize that in the IGIMF theory the gwIMF is not arbitrary but is defined by the metellicity and the SFH of the galaxy and is thus also constrained by the colour of the galaxy.  The main conclusions of these calculations can be summarized as follows.
\begin{itemize}

\item   Based on single stellar population synthesis an age of 6.5 Gyr has been estimated for DF2. But analyzing its stellar population with different adopted starting and truncating times for star formation, we found that DF2's age could be in the range of $4-12.5$ Gyr. This result is based on a color comparison and adopting the reported metallicity of $[Fe/H]=-1$.

\item   Using stellar population synthesis models we show that the number of giant stars can differ by a factor of more than 2 within the observed color uncertainty of DF2. This  may lead to a wrong SBF distance estimation.

\item  A systematically varying gwIMF from top-light in low mass galaxies to top-heavy in massive galaxies as computed by the IGIMF theory, leads to massive galaxies appearing to be located closer, while low mass ones appear further away. However the gwIMF effect is degenerate with the age of the system, as the SBF magnitude is only sensitive to the number of giants, particularly the brightest ones.


\item  Adopting the IGIMF theory, ultra faint dwarfs and ellipticals have  different gwIMFs and number of giants per star in the population. Here we  compare the number of giant stars based on the canonical IMF and the IGIMF in low mass galaxies which reduces the 20 Mpc estimated SBF distance to about 17 Mpc. The combination of the IGIMF,  and intrinsic uncertainties on the SFH (age and metallicity) in DF2 leads to its true distance being possibly as small as  11  Mpc (1$\sigma$ uncertainty in colour) and 7 Mpc (3$\sigma$ uncertainty in colour). This  generally means that using the SBF method, measuring larger distances for galaxies with lower mass could be a natural bias if the true, underlying gwIMF is in fact the IGIMF.



 \end{itemize}

\section*{Acknowledgements}

PK acknowledges support from the Grant Agency of the Czech Republic under grant number 20-21855S. AHZ is supported by an Alexander von Humboldt Foundation postdoctoral research fellowship.

\section*{Data availability}
The data underlying this article are available in the article.

\bsp \label{lastpage} 
\end{document}